\documentclass[12pt]{article}
\usepackage{graphicx}
\begin{document}
\begin{center}
{\LARGE \bf Stellar Structure of Irregular Galaxies. Edge-On Galaxies}

\bigskip

{\large \bf N. A. Tikhonov}

\bigskip

{\em Special Astrophysical Observatory, Russian Academy of Sciences, Nizhnii Arkhyz, Karachai-Cherkessian Republic, 357147 Russia}
\end{center}

\bigskip

 Received July 13, 2005; in final form, November 14, 2005

\bigskip

{\large \bf Abstract} \\

\bigskip

Stellar photometry obtained using the Hubble Space Telescope is used
to study the distributions of the number densities of stars of various
ages in 12 irregular and dwarf spiral galaxies viewed edge-on. Two
subsystems can be distinguished in all the galaxies: a thin disk
comprised of young stars and a thick disk containing a large fraction
of the old stars (primarily red giants) in the system. Variations of
the stellar number densities in the thin and thick disks in the Z
direction perpendicular to the plane of the galaxy follow an exponential
 law. The size of the thin disk corresponds to the visible size of the
galaxy at the $\mu = 25^m$/arcsec$^2$, while the thick disk is a factor
of two to three larger. In addition to a thick disk, the massive irregular
galaxy M82 also has a more extended stellar halo that is flattened at the
galactic poles. The results of our previous study of 12 face-on galaxies
are used together with the new results presented here to construct an
empirical model for the stellar structure of irregular galaxies.

\section{INTRODUCTION}
 Studies of the spatial distribution of the number density of stars
in the stellar subsystems of irregular galaxies requires analyses for a
representative sample of galaxies, including galaxies viewed both edge-on
and face-on. In [1], we presented results for 12 face-on galaxies, and
concluded that these displayed an exponential drop in the radial number
density of old (primarily red-giant) stars, with these stars being
distributed, on average, to distances exceeding the radius of the galaxy
at the $\mu = 25^m$/arcsec$^2$ isophote by a factor of 2.5. However, we ran
into logistical difficulties using the term "thick disk" to describe the
extended structures observed in the face-on irregular galaxies in [1].
The reason is that justification for the use of such a term requires a
simultaneous consideration of the results for both the face-on and edge-on
 galaxies, which was not possible in [1]. We fill this gap here by
presenting our results for 12 edge-on irregular or candidate dwarf-spiral
galaxies.

Several stellar subsystems with different mean ages and
metallicities are usually distinguished in spiral galaxies: the bulge,
thin disk, thick disk, and halo. Although this nomenclature is accepted
by most astronomers, there remain fundamental questions about the
boundaries and sizes of these subsystems. Irregular galaxies have
a simpler stellar morphology. For example, there is no bulge in dwarf
galaxies, and the presence of a halo is debatable. In other words,
only two stellar subsystems are actually observed in irregular galaxies:
the thin and thick disks. We apply this nomenclature by analogy to spiral
galaxies. As in spirals, the thin disk in an irregular galaxy is determined
by the distribution of young stars, while the thick disk is determined by
the distribution of the old stellar population, made up largely by red
giants. The correctness of the use of these terms is demonstrated by the
continuous observational transition in the morphology of the stellar
structure from irregular to spiral galaxies. We thus use the traditional
term "irregular galaxies," while bearing in mind that "irregular" galaxies
are simply low-mass disk galaxies.

\section{GALAXY SAMPLE AND PHOTOMETRY}
 If we limit our consideration to bright supergiants, it is possible to
resolve galaxies into stars using ground-based telescopes. However, the
old stellar population--primarily red giants--is accessible using
ground-based telescopes only in the nearest galaxies, placing severe
limits on possible studies. There are now a large number of deep images
with high angular resolution in the archive of the Hubble Space Telescope
that can be used to study the spatial distributions of various types of
stars in galaxies.

\begin{table}
\caption{Studied galaxies}
\begin{tabular}{llrcrcccl}\hline
\multicolumn{1}{c}{Galaxy}&
\multicolumn{1}{c}{Type}   &
\multicolumn{1}{c}{Apparent}  &
\multicolumn{1}{c}{Distance}&
\multicolumn{2}{c}{Magnitude} &
\multicolumn{1}{l}{i} &
\multicolumn{2}{l}{Disk thickness,kpc}  \\
	    &       &
\multicolumn{1}{l}{size}   &
\multicolumn{1}{l}{Mpc}   &
\multicolumn{1}{l}{$B_T^0$} &
\multicolumn{1}{l}{$M_B$}      &   &
\multicolumn{1}{l}{thin disk}&
\multicolumn{1}{l}{thick disk}\\
\hline
 E381-018   &Irr    &  1.2$\times$ 0.7 & 4.94  &15.8   &$-$13.7: &65 & 0.4 & 1.0  \\
 IC1574     &IB(s)m &  2.1$\times$ 0.8 & 5.42  &15.1   &$-$14.8  &90 & 0.6 & 1.8  \\
 IC3104     &IB(s)m &  3.8$\times$ 1.8 & 2.30  &13.6   &$-$15.7  &82 & 0.3 & 1.6: \\
 NGC1560    &SA(s)d & 11.7$\times$ 1.9 & 3.48  &12.2   &$-$18.1  &90 & 0.6 & 1.7: \\
 PGC1641    &dS0/Im &  1.1$\times$ 0.7 & 2.07  &15.6   &$-$11.8: &90 & 0.2 & 0.5  \\
 PGC9962    &Scd    &  7.2$\times$ 0.8 & 4.72  &13.2   &$-$16.3  &90 & 0.6 & 1.4  \\
 PGC39032   &IBm    &  1.4$\times$ 0.6 & 3.19  &15.2   &$-$13.4  &90 & 0.4 & 1.0  \\
 UGC1281    &Sdm    &  5.8$\times$ 0.7 & 5.11  &12.9   &$-$16.9  &90 & 0.4 & 1.2  \\
 UGCA442    &SB(s)m &  6.4$\times$ 0.9 & 4.61  &13.6   &$-$15.6  &90 & 0.8 & 2.0  \\
 7Zw403     &Pec    &  1.4$\times$ 0.8 & 4.35  &14.5   &$-$14.6: &54 & 0.6 & 1.7  \\
 M82        &I0     & 11.2$\times$ 4.3 & 3.90  & 9.3   &$-$19.3: &87 & 1.5:& 2.9  \\
\end{tabular}
\end{table}

As in our earlier paper [1], the criteria for selecting
the galaxies for the study were their sizes (which are substantial for the
HST WFPC2 camera) and
rotational rates ($V_r <$ 100 km s$^{-1}$; this indirectly determines
the galaxy
type). Since the inclinations of most irregular galaxies to the line of
sight are known only with large uncertainty, we selected galaxies with
probable large axial ratios $a/b$, assuming that precisely such galaxies
will likely be inclined to the line of sight by angles close to 90.
Using the list of nearby galaxies in the Local Group [2], we determined
that different types of galaxies have different mean axial ratios $a/b$.
The mean axial ratios for galaxies of types T = 7, 8, 9, and 10 are
$a/b$ = 5.3, 2.9, 2.1, and 1.8. We can clearly see a tendency for decreasing
a/b in the transition from obviously spiral to irregular galaxies. Among
225 nearby irregular galaxies (T = 10) with luminosities below $M_B = -14$,
there are only seven with axial ratios $3 < a/b < 5$. Two of these are
located in regions of strong absorption in the Milky Way, and their
morphologies are not clear. Among the remaining five galaxies, only
UGC 5186 has a high axial ratio, $a/b$ = 4.3, while the other galaxies
have $3.0 < a/b < 3.4$. This suggests that the most flattened galaxies
are more likely to be spirals than irregulars according to a Hubble
classification, with the axial ratios of irregular galaxies being,
as a rule, $a/b < 3$.

Understanding that applying too strict criteria
for the selection of the galaxies will lead to a very small number
of objects in the sample, and bearing in mind possible inaccuracies
in the classification of galaxies, we left the two probable dwarf
spiral galaxies NGC~1560 and UGC~A442 in our list. The three
galaxies NGC~1560, UGC~1281, and PGC~9962 have rotational rates that
slightly exceed our adopted criterion (100 km s$^{-1}$), however we left these
galaxies in our list, in view of the absence of evident spiral structure
in HST images.

A list of the galaxies studied is given in the table,
which presents for each galaxy the visible size and morphological type
from the NASA Extragalactic Database (NED), the inclination from the
LEDA database {\em (http://www-obs.univ-lion1.fr/hypercat/ search.html/)},
the integrated magnitude $B_T$ from the RC3 catalog, and the distance,
absolute magnitude, and total thickness of the thin and thick disks
in the $Z$ direction derived by us. The three galaxies ESO~381-018,
UGC~7242, and 7Zw403 have LEDA inclinations 60--65$^\circ$, but, as we indicated
above, the uncertainties of such data can be quite high. For example,
the bright part of 7Zw403 used to determine the inclination is compact.
However, the thick disk of this galaxy, which is comprised of red
giants and is visible only in HST images, is appreciably elongated,
suggesting a high inclination of the galaxy to the line of sight.

We reduced the HST WFPC2 and ACS/WFC images using the MIDAS package.
We did not perform the preliminary reduction, since this was done
automatically when we requested the images from the HST archive.
Cosmic-ray traces were removed using the standard FILTER/COSMIC program
in MIDAS. We carried out photometry of the stars in the images using
the DAOPHOT II [3] and HSTphot [4] programs.

\begin{figure}[h]
\centerline{
\includegraphics[bb=149 388 438 667,width=9.cm,angle=0,clip]{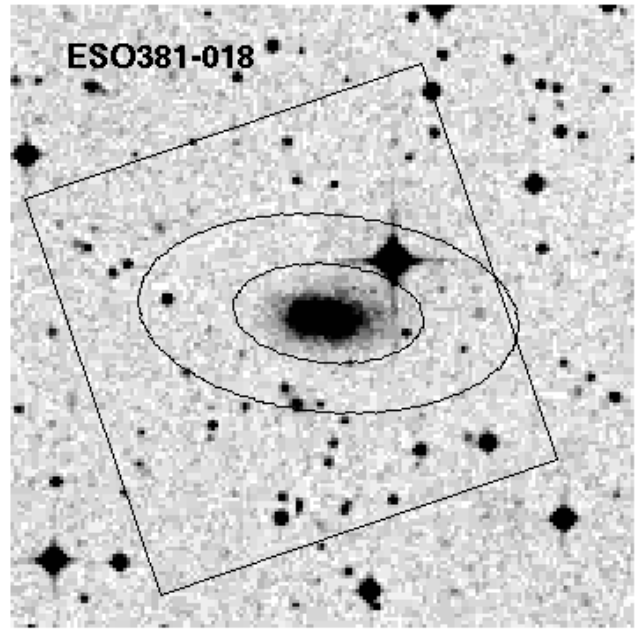} }
\caption{5${\prime}\times5^{\prime}$ image of the galaxy
ESO 381-018 in a frame of the DSS survey. The square marks
the region of the ACS/WFC image taken with the Hubble Space
Telescope. The ellipses denote the derived boundaries of the
thin and thick disks.}
\end{figure}

\begin{figure}[h]
\centerline{
\includegraphics[bb=122 134 348 375,width=9.cm,angle=-90,clip]{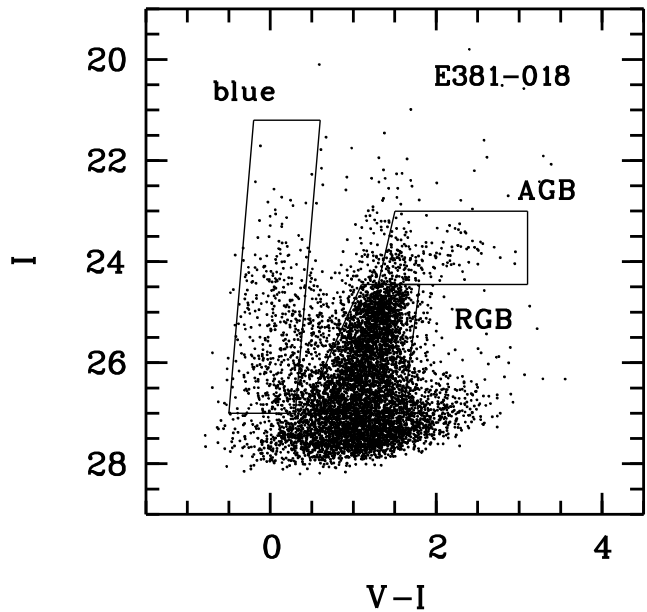}}
\caption{Herzsprung--Russell diagram (CMD) for the stars in the HST
images after photometry using DAOPHOT II. The solid curves show the
boundaries of regions within which stars of various types (blue,
AGB, RGB) were chosen when determining their spatial distributions
in the galaxy.}
\end{figure}

HSTphot yields more accurate results, as can be seen from the decrease in
the width of the red-giant and blue-giant branches in color--magnitude
diagrams, but DAOPHOT II is able to photometrize weaker stars than HSTphot,
which can sometimes be of crucial importance.

We applied the recommendations
given in [5, 6] when carrying out the photometry and translating the
instrumental magnitudes for the DAOPHOT II WFPC2 images to the standard
VI Kron--Cousins system. The translation of the instrumental magnitudes
for the ACS/WFC camera to the standard VI Kron--Cousins system was based
on calibration relations obtained for stars in the irregular galaxy IC10,
for which we performed photometry for the same stars using the ACS/WFC and
 WFPC2 images.

The photometry yielded tables with the coordinates and
magnitudes of the stars, together with accompanying parameters enabling
estimation of the photometric accuracy and deviations of the profile of
a photometrized object from the standard stellar profile, making it
possible to remove diffuse objects (distant galaxies) and pseudo-stars
arising due to residual cosmic-ray traces from the final list of stars.

\section{DISTRIBUTION OF THE NUMBER DENSITY OF STARS IN THE Z DIRECTION}
 Using one galaxy as an example, let us consider the procedure used for
our measurements. Figure 1 presents
5$^{\prime}\times5^{\prime}$ images of the Digital Sky Survey (DSS) for the Southern galaxy
ESO~381-018. The square marks the region imaged by the HST ACS/WFC.
The ellipses indicate the boundaries of the distributions of young
stars (thin disk) and old stars (thick disk) found by us. Figure 2
presents the results of stellar photometry of the ACS/WFC image in
the form of a color--magnitude diagram (CMD). This diagram is an
ordinary Hertzsprung--Russell diagram for irregular galaxies.
The red-giant branch (RGB) and blue-supergiant branch (blue) are
clearly distinguished. The red-supergiant branch and asymptotic
giant branch (AGB) are less clearly visible due to their smaller
populations. The regions whose stars were used to calculate the
stellar number densities are indicated by lines. Figure 3 presents
the distribution in the galaxy of the blue and red stars distinguished
in this way. As in the face-on galaxies, the blue stars are concentrated
toward the galactic center, while the red stars form a more extended
subsystem. The parallel lines in Fig. 3 indicate bands within which
the stellar number densities were calculated; the results are presented
in Fig. 4, which shows the distributions of stars along both the major
and minor ($Z$) axes of the galaxy. We can see that the size of the galaxy
corresponding to the distribution of red giants is 4 kpc$\times$ 2 kpc or
3.0$^{\prime}\times1.5^{\prime}$ , which exceeds the size of the galaxy given in the NED
database by a factor of two. The blue supergiants form a subsystem
0.8 kpc × 0.4 kpc in size, while the AGB stars form a subsystem with
an intermediate size, which we do
not present in view of the small number of AGB stars and associated
large fluctuations in the stellar number density.

\begin{figure}
\includegraphics[bb=123 44 373 463,width=9.cm,angle=-90,clip]{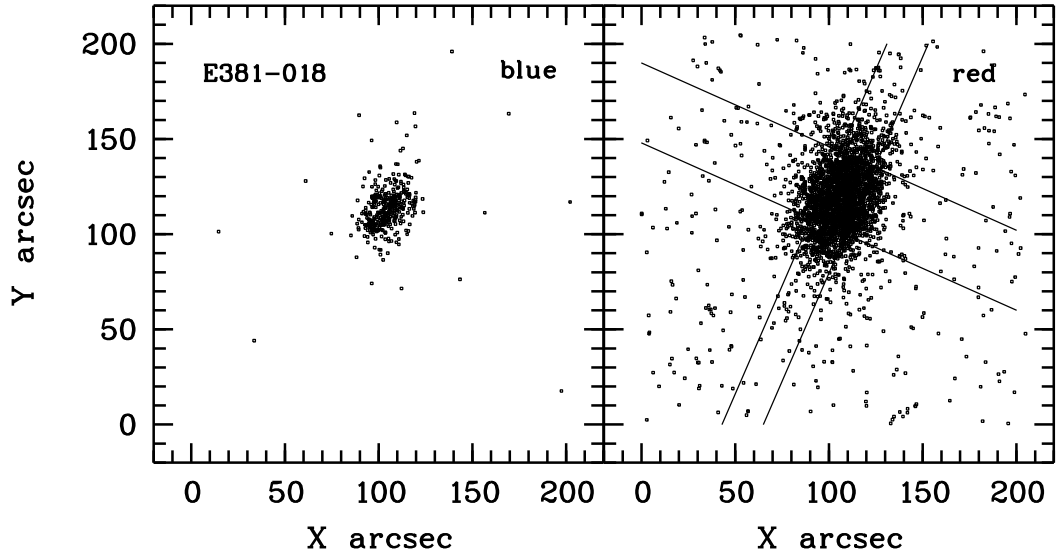}
\caption{Distribution of blue stars (left) and red giants (right) in
the HST frame for the galaxy ESO 381-018. The parallel lines denote bands
within which the stellar number density was determined along the major axis
and the $Z$ axis of the galaxy.}
\end{figure}

\begin{figure}
\includegraphics[bb=117 55 370 497,width=9.cm,angle=-90,clip]{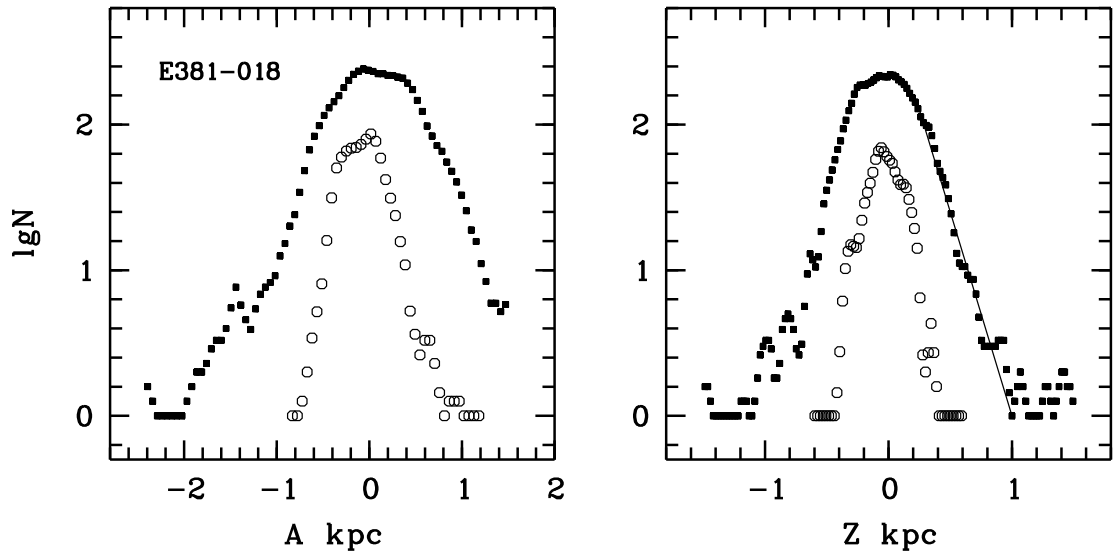}
\caption{Distribution of the number density of blue stars (hollow squares)
and red giants (filled squares) along the major axis (left) and along
the $Z$ axis (right) for the galaxy ESO 381-018.}
\end{figure}

Since a red-giant
branch can be distinguished in each galaxy, we used the "Tip of the
Red-Giant Branch" method to calculate the distance to each galaxy [7].
To simplify the calculations, although this yields a somewhat lower
accuracy, we took the absolute magnitude of stars at the tip of the
red-giant branch to be $M_I = -4.05$ for all galaxies. This approach can
worsen the accuracy in the distance modulus by up to 0.05$^m$, but this
is not important for the problem at hand. We used the extinctions of [8]
when calculating each distance modulus.

We measured the number density of
stars for the other galaxies in our list in the same way as for
ESO~381-018; the results are shown in Fig. 5. In spite of some differences
in the galactic morphologies, we can see that, in each of the 11 galaxies,
the red giants form a subsystem that is a factor of two to three larger
than the subsystem of young stars. We can see that the red giants display
an exponential decrease in the number density of stars in the $Z$ direction.
The deviation from exponential behavior in the fall-off of the number
density of red giants for NGC 1560 has a technical origin--a decrease
in the area for the stellar counts due to an unfortunate positioning
of the center of the WFPC2 image relative to the galaxy. The largest
ratio of the sizes of the thick and thin disks is displayed by IC 3104
(type IB(s)m). However, the galaxy IC 1574, which has the same type,
displays a completely normal ratio of these sizes.

\begin{figure}
\includegraphics[bb=67 210 468 826,width=12.cm,angle=0,clip]{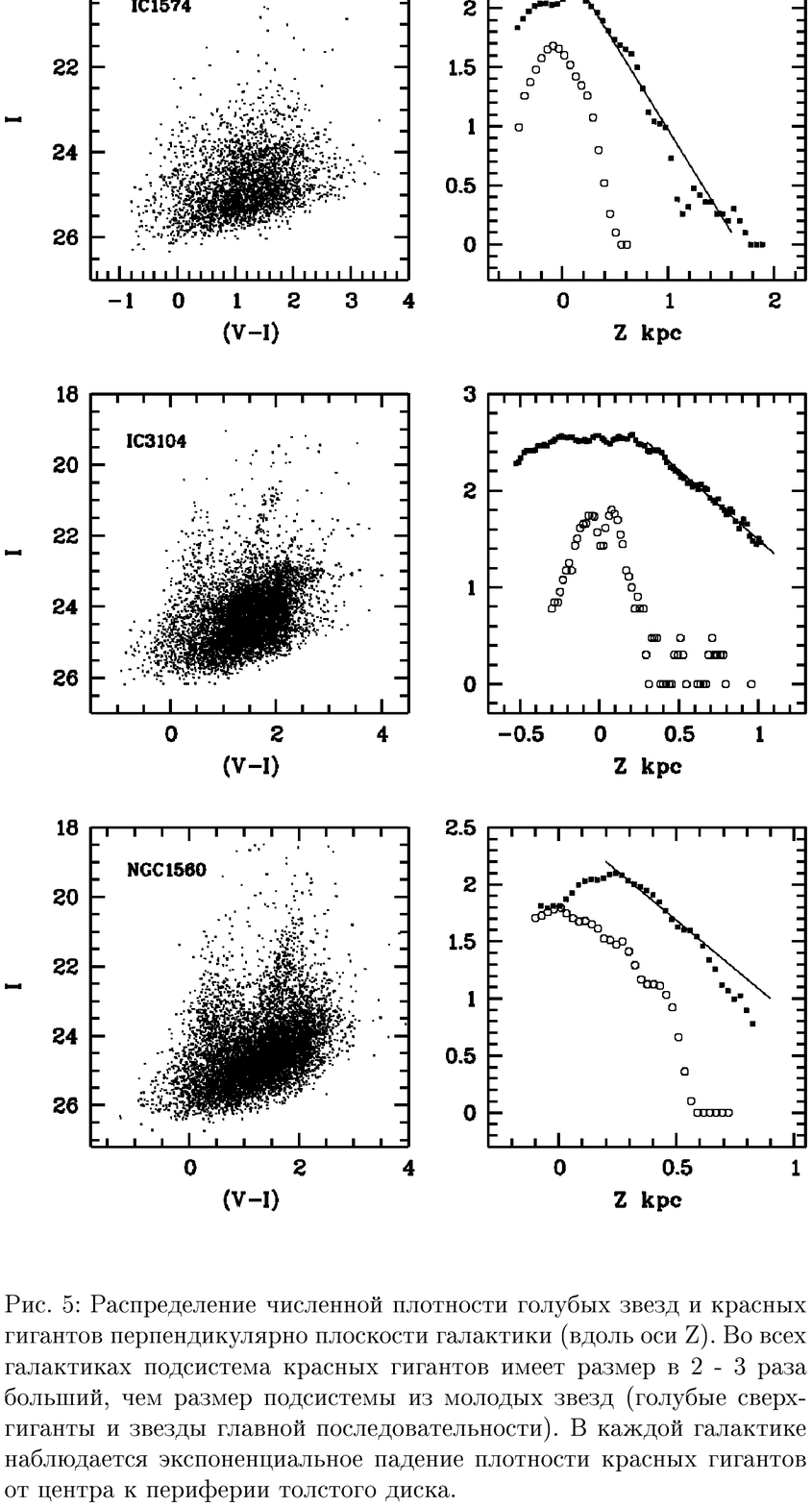}
\caption{Hertzsprung--Russell diagrams and distributions of the number
densities of blue stars (hollow squares) and red giants (filled squares)
perpendicular to the galactic plane (in the $Z$ direction) for the studied
galaxies. In all cases, the subsystem of red giants is a factor of two to
three larger in size than the subsystem of young stars (blue supergiants
and main-sequence stars). In each galaxy, we observe an exponential drop
in the number density of red giants from the center to the periphery of
the thick disk.}
\end{figure}

\setcounter{figure}{4}
\begin{figure}
\includegraphics[bb=67 210 468 826,width=12.cm,angle=0,clip]{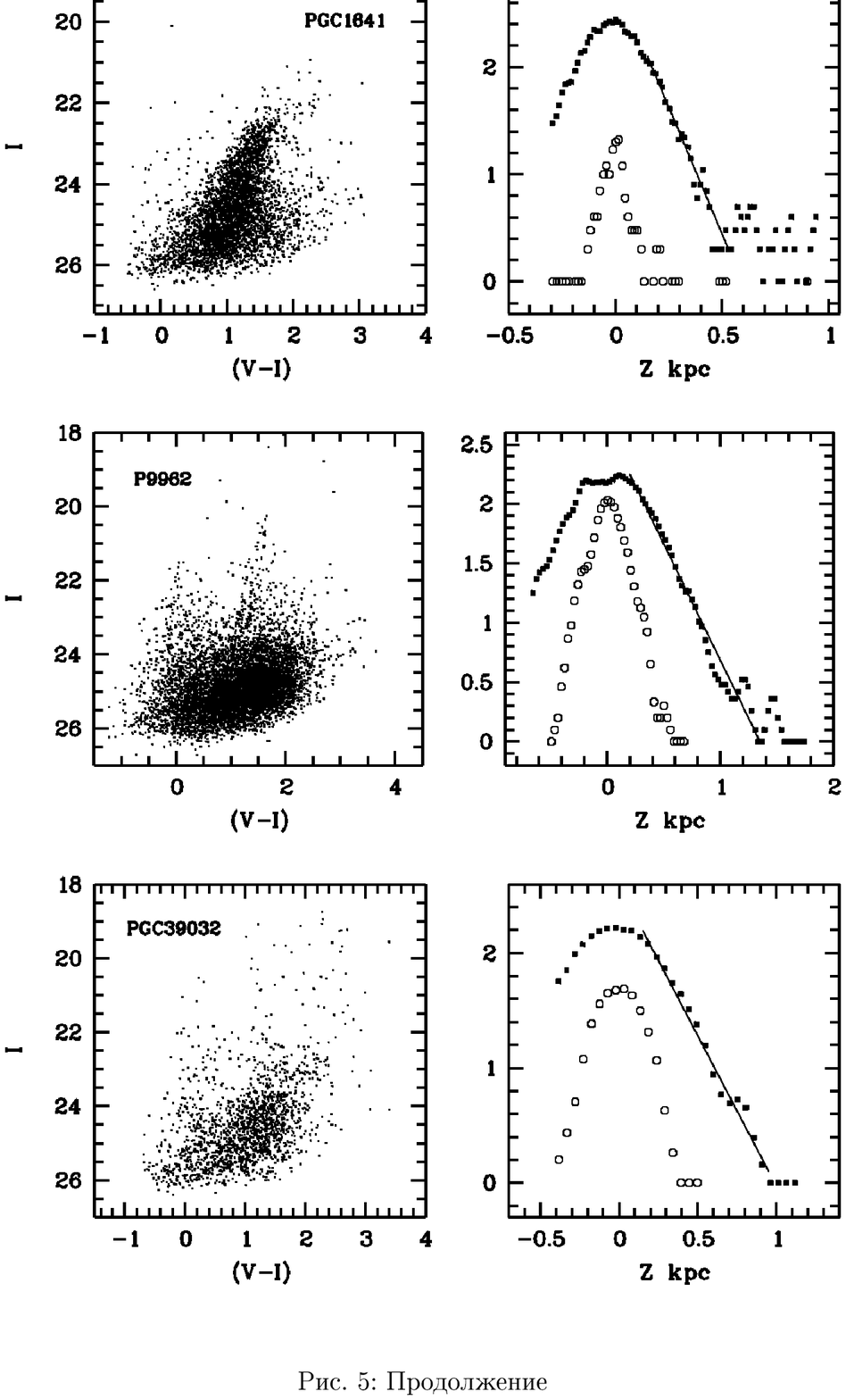}
\caption{Contd.}
\end{figure}

\setcounter{figure}{4}
\begin{figure}
\includegraphics[bb=67 210 468 826,width=12.cm,angle=0,clip]{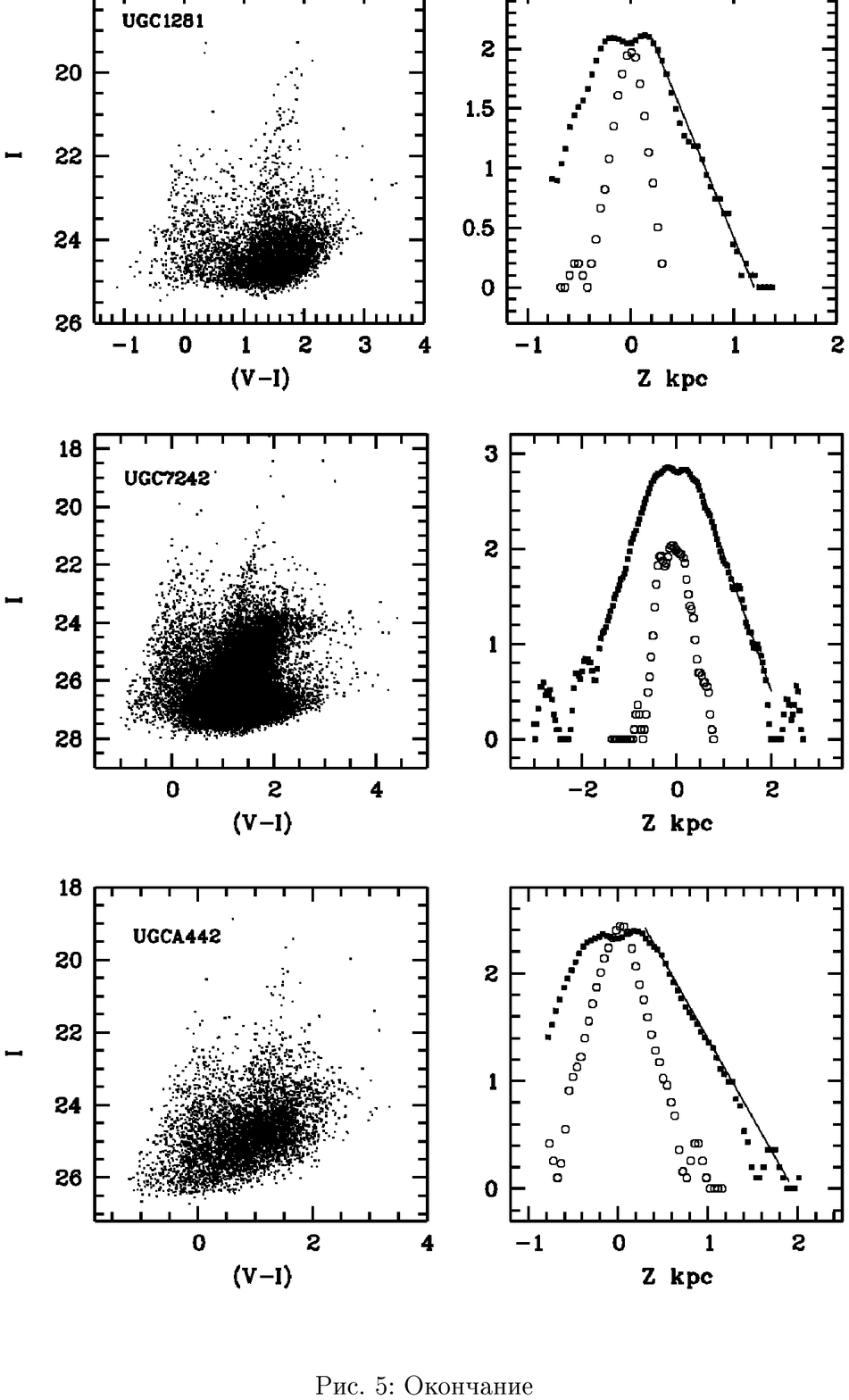}
\caption{Contd.}
\end{figure}

An elongated disk of red giants can be seen around the blue compact galaxy
with intense central star formation 7Zw403, which has a typical ratio of
thick-to-thin disk sizes, confirming that it is an irregular galaxy
undergoing a burst of star formation. The galaxy M82 was comparatively
recently considered to be a prototype Irr II irregular galaxy. It has now
been firmly established that the anomalous colors of the galaxy are a
consequence of absorption in extended gas--dust clouds. Although M82
has a high mass, no spiral structure is observed. The edges of the ACS/WFC
image of this galaxy (Fig. 6) coincide with the boundary of the thick disk,
beyond which a halo is clearly detected. M82 is one of only a few  irregular
galaxies in which an extended halo that is flattened at the galactic poles
is observed, in addition to the thick disk (Fig. 6). Other irregular galaxies
with halos include IC10 [9] and NGC 3077.

We expect the galaxy NGC 1560 to
 have a halo based on its brightness and morphology, but were not able to
find images of regions far from the center of this galaxy in the HST
archive. The young stars in the thin disk also display an exponential
fall-off in their number density in the $Z$ direction. This is most clearly
visible in galaxies with a large number of young stars, since there can
be substantial statistical fluctuations in the stellar number density if
the number of stars is small.

As was noted above, not all of the galaxies considered have formal
inclinations to the line of sight of 90. The results presented in
the table show that the mean total sizes for the thin and thick disks
in the Z direction are 1.2 and 3.2 kpc; i.e., they are not very thin.
The possible importance of corrections to the galaxy inclinations can
be estimated from the following example. For a galaxy with a ratio of
its thickness to its diameter of 1 : 3, which approximately corresponds
to the results for the irregular galaxies in the table, the correction
to the size of the disk due to the galaxy having an inclination
differing by 30$^\circ$ from the required value of 90$\circ$ leads to an increase
in the calculated size of the disk by only 10\%, which is not important,
given the other possible sources of measurement error. In other words,
the $Z$ sizes of the thin and thick disks in the table correspond to
their
true sizes, with only small corrections for the effects of inclination.

\section{THE STRUCTURE OF IRREGULAR GALAXIES}
 Given the randomness of our selection of the studied galaxies, it is
very likely that other irregular galaxies have a similar structure: a
thin and thick disk, with a halo as well in massive galaxies. The thin
disk is defined by the boundary in the distribution of young stars.
Naturally, the lower the mass of the galaxy, the larger the
fluctuations in the shape of the thin disk that can be introduced
by bright star-forming regions. Although the mean distribution of
young stars follows an exponential (or close to exponential) law
in the $Z$ direction, deviations from such a law
are possible in some galaxies due to the presence of regions of star
formation near the edge of a disk.

Since the thick disks of irregular
galaxies consist of old stars, their distributions of the number density
of stars should display monotonic properties. To all appearances, only
external gravitational influences can distort the observed exponendial
variation in the stellar number density for the thick disk in the $Z$
direction, as is observed, for example, in the spiral galaxy NGC 4631
[10]. The size of the thick disk in the $Z$ direction is a factor of two
to three larger than the size of the thin disk, which corresponds within
the uncertainties to the ratio of the radii of these disks for the face-on
galaxies.

The relative dimensions of the halos of irregular galaxies remain
uncertain due to the small-number statistics for our sample. A comparison
of the stellar structures for
edge-on irregular galaxies and spiral galaxies [10--12] indicates
that the two types of galaxies are morphologically similar in terms
of the relative sizes of their thin and thick disks. This suggests
that the relative halo dimensions for irregular galaxies may likewise
obey the same relations we found for spiral galaxies. Thus, the halos
should be flattened at the galactic poles and extend in the Z
direction to distances about twice as far as the thick disks.

The question of whether the cross sections of disk galaxies are
elliptical or oval remains unresolved due to insufficient data. The
difference between these shapes is not large enough to be able to
distinguish between them in the presence of large fluctuations in
the surface density of stars at the edge of the galaxy. Moreover,
it is not ruled out that both elliptical and oval cross sectons could
be present in a galaxy.

 \begin{figure}[h]

\vspace*{-2cm}
\hspace*{1.0cm}
\includegraphics[height=7.cm,angle=0,clip]{fig_6a.ps}

\vspace*{-7.cm}
\hspace*{8cm}
\includegraphics[height=7.cm,angle=-90,bb=91 37 464 400,clip]{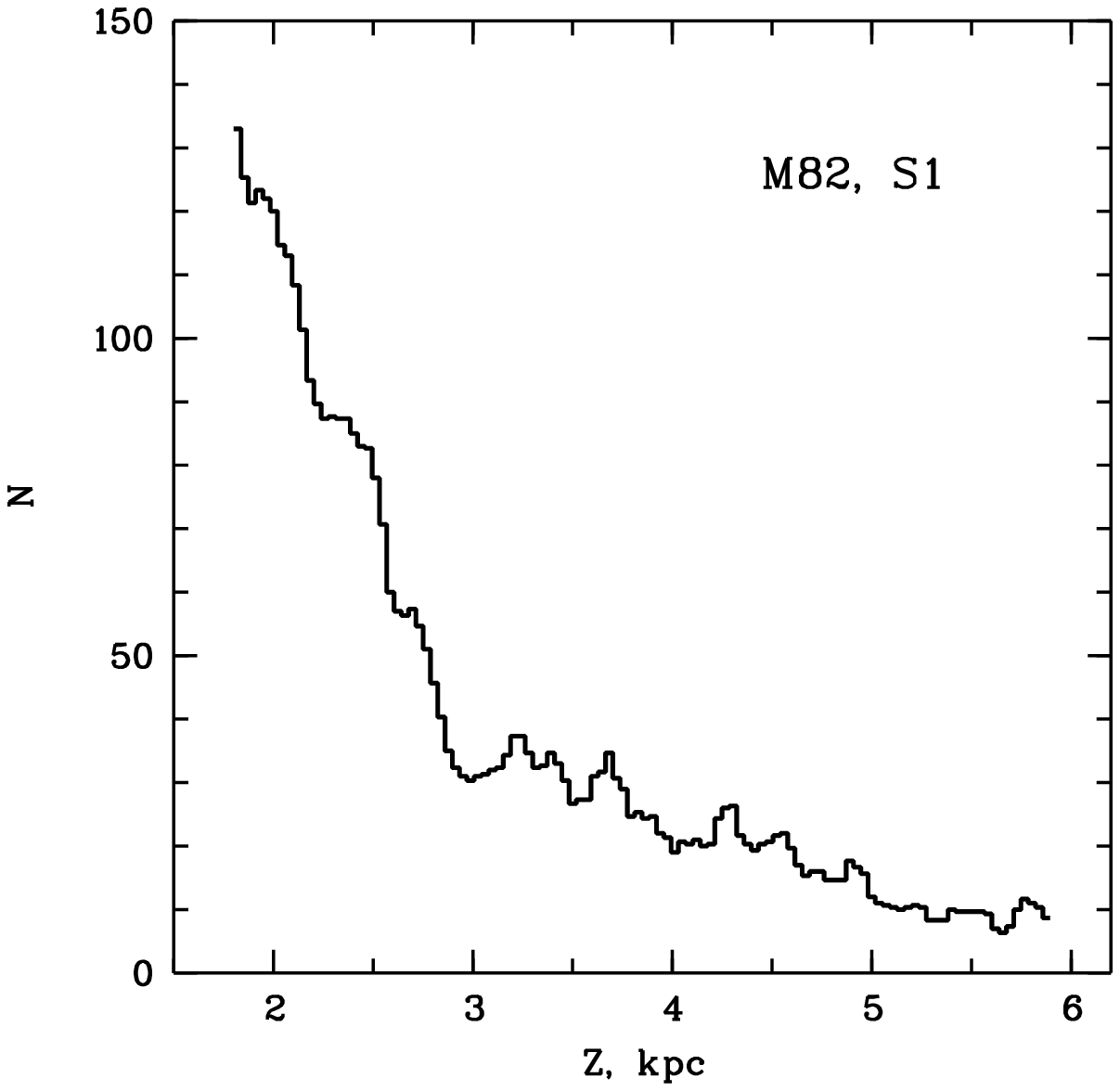}

\caption{Left: image of 82 from the DSS with the ACS/WFC field and boundary
between the thick disk and halo indicated. Right: distribution of the number
density of red giants in 82 in the $Z$ direction. There is a sharp variation
in the gradient of the number density of red giants at the boundary between
the thick disk and halo, which we take to represent the edge of the thick
disk.}
\end{figure}

\begin{figure}
\centerline{
\includegraphics[height=7.cm,angle=0,clip]{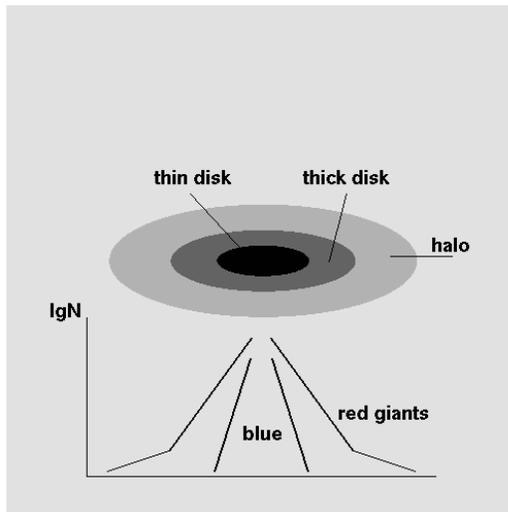}}
\caption{Empirical model for the stellar structure of an edge-on irregular
galaxy consisting of a thick disk, thin disk, and halo. The halo is absent
in dwarf galaxies. The radial behavior of the number densities of blue stars
and red giants is shown schematically in the lower part of the figure. The
distributionof stars in the Z directionhas a very similar appearance.The
relativesizes of the disks and halo in the figure correspond to the mean
values obtained for the 24 galaxies studied.}
\end{figure}

Based on our studies of 24 face-on and edge-on galaxies, we have constructed
a simple empirical model for the stellar structures of irregular galaxies
(Fig. 7). We cannot indicate an absolute scale in Fig. 7 because the sizes
and degree of flattening vary from galaxy to galaxy, but the mean ratio of
the sizes of the thick and thin disks is a fairly stable quantity, equal
to 2.5$\pm$0.5.

\section{DISCUSSION AND CONCLUSIONS}
In our studies of irregular galaxies, we have used the young and old
stellar populations to construct the stellar distributions in the thin
and thick disks. The intermediate-age population (AGB stars) have remained
unused for several reasons: (a) the small number of AGB stars in the galaxy
peripheries makes any conclusions very uncertain, and (b) there are both
relatively young, high-mass stars and comparatively old, low-mass stars
within the region in Fig. 2 occupied by AGB stars. It is not possible
to separate them out based on our photometry. Thus, results for the
distribution of AGB stars will depend on ratios of the numbers of stars
of various ages, but should, on average, correspond to the distribution
of intermediateage stars (with ages between those for blue supergiants
and red giants). This is indeed observed in the studied galaxies.

Our
empirical model for the stellar structures of irregular galaxies is
consistent with the results of other studies to date [13--18]. The
only exception is the dwarf galaxy Leo A, for which both a thick disk
and halo have been detected, in contradiction with our suggestion that
halos should be observed for massive galaxies. Since the method of
fitting ellipses was used to describe the distribution of the stellar
number density for this galaxy, which differs from our method of using
narrow bands or sectors, it may be that the appearance of the halo of
this galaxy is associated with the use of different search methods. Indeed,
although ellipse-fitting yields deeper results for star counts, it can lead
to spurious variations in the gradient of
the radial distribution when the isophote eccentricity varies with
distance from the galactic center, which can easily be misinterpreted
as the presence of a halo.

In summary, our studies of the stellar
populations in 24 face-on and edge-on galaxies have yielded the following
conclusions.

(1) The distributions of stars of various ages in the radial
and $Z$ directions follow exponential (or close to exponential) laws that
correspond to the disk component of the galaxy.

(2) Young and old stars
have different gradients for the fall-off of number density from the center
to the edge of the galaxy.

(3) Young stars for the thin disk, whose size
corresponds approximately to the visible size of the galaxy, while the old
stars (red giants) form the thick disk, whose size is a factor of two to
three larger than thin.

(4) A halo made up primarily of red giants is also
observed in massive irregular galaxies, beyond the thick disk.

(5) Based
on the measured number density of red giants, irregular galaxies have
fairly sharp boundaries. The implied sizes of the galaxies exceed the
sizes in catalogs such as the NASA Extragalactic Database by factors of
two to three.

The derived characteristics of irregular galaxies indicate
that the erroneous term "irregular" is no more than a remnant of earlier
work, since all these galaxies have a regular axially symmetrical
structure; i.e., they are all low-mass disk galaxies. Moreover, there
is no demarcation boundary between the morphologies of the stellar
structures of irregular and spiral galaxies. They are all disk galaxies,
and differ only in their masses, which leads over the time scale for
their evolution to differing forms and appearances, such as evidence
spiral structure.

{\large \bf ACKNOWLEDGMENTS}
This work was supported by the Russian Foundation for Basic Research
(project no. 03--02--16344), and made use of the NASA/IPAC Extragalactic
Database.

\bigskip

{\bf REFERENCES} \\
1. N. Tikhonov, Astron. Zh. 82, 1 (2005) [Astron. Rep. 49, 501 (2005)].

2. I. Karachentsev, V. Karachentseva, W. Huchtmeier, et al., Astron. J. 127, 2031 (2004).

3. P. Stetson, User Manual for DAOPHOT II (1994).

4. A. Dolphin, Publ. Astron. Soc. Pac. 112, 1383 (2000).

5. J. Holtzman, J. Hester, S. Casertano, et al., Publ. Astron. Soc. Pac. 107, 156 (1995).

6. J. Holtzman, C. Burrows, S. Casertano, et al., Publ. Astron. Soc. Pac. 107, 1065 (1995).

7. M. Lee, W. Freedman, and B. Madore, Astrophys. J. 417, 553 (1993).

8. D. Schlegel, D. Finkbeiner, and M. Davis, Astrophys. J. 500, 523 (1998).

9. I. Drozdovsky, N. Tikhonov, and R. Schulte-Ladbeck, Rev. Mex. Astron. Astrofis. 17, 46 (2003).

10. N. Tikhonov, O. Galazutdinova, and I. Drozdovsky, Mon. Not. R. Astron. Soc. (2005, in press).

11. N. Tikhonov, O. Galazutdinova, and I. Drozdovsky, Astron. Astrophys. 431, 127 (2005).

12. N. Tikhonov and O. Galazutdinova, Astrophys. 48, 221 (2005); astro-ph/0503235 (2005).

13. D. Minniti and A. Zijlstra, Astrophys. J. Lett. 467, L13 (1996).

14. D. Minniti, A. Zijlstra, and M. Alonso, Astron. J. 117, 881 (1999).

15. S. Demers, P. Battinelli, and B. Letarte, Astron. Astrophys. 410, 795 (2003).

16. B. Letarte, S. Demers, P. Battinelli, and W. Kunkel, Astron. J. 123, 832 (2002).

17. R. Lynds, E. Tolstoy, E. J. O'Neil, and D. Hunter, Astron. J. 116, 146 (1998).

18. A. Aparicio and N. Tikhonov, Astron. J. 119, 2183 (2000).

19. V. Vansevicius, N. Arimoto, T. Hasegava, et al., Astrophys. J. Lett. 611, L93 (2004).

\end{document}